\documentclass[%
 aip,
 amsmath,amssymb,
 reprint,%
]{revtex4-1}
\usepackage{hyperref} 
\usepackage{xcolor}
\usepackage{graphicx}
\usepackage{dcolumn}
\usepackage{bm}

\usepackage[utf8]{inputenc}
\usepackage[T1]{fontenc}
\usepackage{mathptmx}
\usepackage{etoolbox}

\makeatletter
\def\@email#1#2{%
 \endgroup
 \patchcmd{\titleblock@produce}
  {\frontmatter@RRAPformat}
  {\frontmatter@RRAPformat{\produce@RRAP{*#1\href{mailto:#2}{#2}}}\frontmatter@RRAPformat}
  {}{}
}%
\makeatother
\draft 
\begin{document}
\preprint{AIP/123-QED}

\title{Future Research Perspective on the Interfacial Physics of Non-Invasive Glaucoma Testing in Pathogen Transmission from the Eyes}



\author{Durbar Roy}
\author{Saptarshi Basu$^{*}$}

\email[Corresponding author:]{sbasu@iisc.ac.in}
\affiliation{Department of Mechanical Engineering, Indian Institute of Science, Bengaluru, KA 560012, India}


\date{\today}

\begin{abstract}
Non-contact Tonometry (NCT) is a non-invasive ophthalmologic technique to measure intraocular pressure (IOP) using an air puff for routine glaucoma testing. Although IOP measurement using NCT has been perfected over many years, various phenomenological aspects of interfacial physics, fluid structure interaction, waves on corneal surface, and pathogen transmission routes to name a few are inherently unexplored.
Research investigating the interdisciplinary physics of the ocular biointerface and of the NCT procedure is sparse and hence remains to be explored in sufficient depth. In this perspective piece, we introduce NCT and propose future research prospects that can be undertaken for a better understanding of the various hydrodynamic processes that occur during NCT from a pathogen transmission viewpoint. 
In particular, the research directions include the characterization and measurement of the incoming air puff, understanding the complex fluid-solid interactions occurring between the air puff and the human eye for measuring IOP, investigating the various waves that form and travel; tear film breakup and subsequent droplet formation mechanisms at various spatiotemporal length scales. Further, from ocular disease transmission perspective, the disintegration of the tear film into droplets and aerosols poses a potential pathogen transmission route during NCT for pathogens residing in nasolacrimal and nasopharynx pathways. Adequate precautions by opthalmologist and medical practioners are therefore necessary  to conduct the IOP measurements in a clinically safer way to prevent the risk associated with pathogen transmission from ocular diseases like conjunctivitis, keratitis and COVID-19 during the NCT procedure.
\end{abstract}


\maketitle 

\section{Introduction}
 Non-contact tonometry (NCT) \cite{kniestedt2008tonometry} is a widely used technique for measuring intraocular pressure (IOP) \cite{shiose1990intraocular,stamper2011history}; a key ophthalmologic diagnostic indicator for various ocular conditions, including glaucoma \cite{quigley1993open,quigley1996number,thylefors1994global,lee2005glaucoma}. Early detection of glaucoma helps ophthalmologists treat conditions originating from high IOP. Excess and irregular IOP can cause stress in the optic nerve causing damage to nerve fibers and resulting in the formation of peripheral blind spots, tunnel vision, and permanent blindness \cite{albert2008albert}. Intraocular pressure (IOP) refers to the fluid pressure inside the anterior chamber of the eye, which is maintained by the balance of inflow and outflow of aqueous humor; a clear watery fluid that circulates within the eye \cite{shiose1990intraocular,stamper2011history}. IOP is an important parameter in evaluating ocular health. It is used as a screening test for several conditions that can cause irreversible damage to the optic nerve and result in vision loss \cite{bourne2013causes}. 
 IOP is generally measured as a gauge pressure (i.e., pressure above the atmospheric pressure) in mm of Hg. Given the importance of such non-invasive procedures like NCT in ocular health diagnostics, it is important to investigate the safety of various ophthalmologic procedures from a clinical and mechanistic perspective. Previous studies shows \cite{roy2021fluid,britt1991microaerosol} micro-aerosols and drop formation can occur from corneal tear film during NCT procedure. Such drops and aerosols generated can lead to new pathogen transmission routes for microorganisms present in nasolacrimal and nasopharynx pathways during non-invasive eye procedures. Pathogen present in human tears can spread from drops and aerosols originating due to tear film destabilization due to a complex hydrodynamic interaction that occurs between the corneal tear film and the incoming air-puff as was shown in our previous studies \cite{roy2021fluid,shetty2020quantitative}. Further in-depth research into various mechanistic processes are hence required to understand pathogen transmission during NCT in a quantitative perspective. This article highlights some of the future research prospects that can be taken in the future to expand our understanding of NCT and pathogen transmission during NCT.

 Tonometry in general is a diagnostic test to measure the intraocular pressure (IOP) \cite{kniestedt2008tonometry,cook2012systematic} and is an essential tool in diagnosing and managing various ocular conditions. IOP measurements are performed using specialized instruments called tonometers, which can be subdivided into two major classes, contact and non-contact \cite{shields1980non,tonnu2005influence}. Contact tonometry involves touching the cornea with a small, flat probe to measure the force needed to flatten a specific corneal area. The most commonly used contact tonometer and the gold standard is the Goldmann applanation tonometer \cite{kaufmann2004comparison,moses1958goldmann}. The test requires a topical anesthetic (local anesthesia) to numb the eye's surface before gently applying the probe to the cornea \cite{brusini2006comparison}. Non-contact tonometry, on the other hand, uses a puff of air to measure IOP. The non-contact test is performed using a device called a non-contact tonometer or air-puff tonometer \cite{farhood2012comparative}. The patient is seated in front of the device, and a puff of air is directed at the cornea. The device then measures the IOP based on the corneal 
deflection in response to the air puff. While both contact and non-contact tonometry are accurate methods of measuring IOP, there are some differences between the two techniques \cite{farhood2012comparative}. Contact tonometry is considered the gold standard for IOP measurement, as it provides more precise and reliable readings. However, it requires anesthetic drops and may be uncomfortable for some patients. Non-contact tonometry is a more comfortable alternative and helpful in screening patients who cannot tolerate contact tonometry. It is important to note that IOP readings can vary, just like blood pressure throughout the day, and a single measurement may not be sufficient to diagnose a condition such as glaucoma \cite{bagga2009intraocular,david1992diurnal}.
 Hence, a series of IOP measurements over time may be necessary to accurately assess changes in IOP and determine the best course of treatment. For repeated temporal measurements of IOP, non-contact methods are more practical and comfortable for the patients. 
Further, in a general hospital setting non contact mode is more efficient and reliable to handle a large volume of patients.
 NCT is a safe and non-invasive procedure that can be performed in a doctor's office or clinic. It is a crucial part of routine eye exams, particularly for patients at high risk for developing glaucoma. Early detection and management of elevated IOP can help prevent vision loss and preserve ocular health. Contact and non-contact tonometry are accurate methods of measuring IOP, and the choice of technique depends on the individual patient's needs and preferences. 
 \begin{figure*}
  \centerline{\includegraphics[scale=1.0]{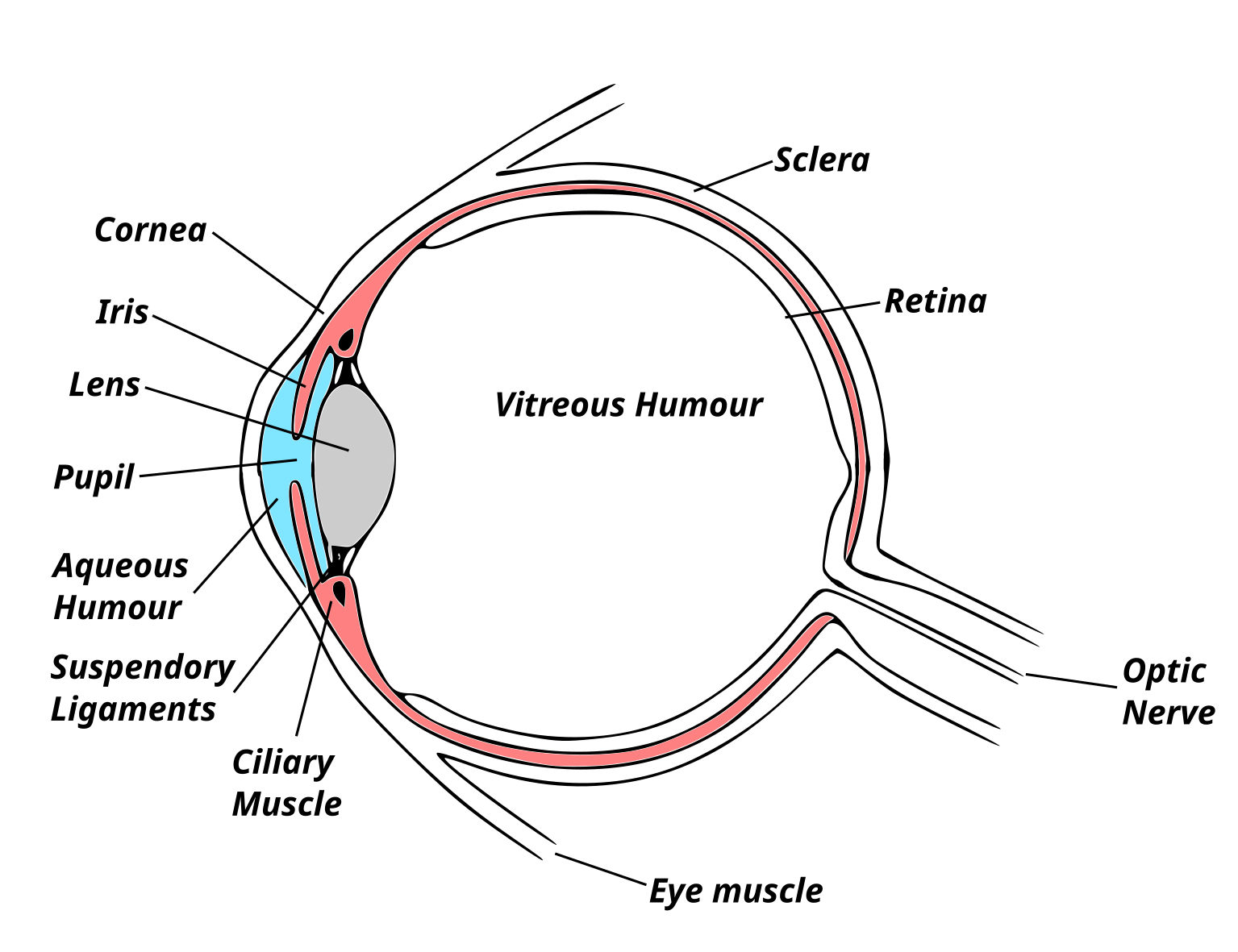}}
  \caption{Schematic showing the cross sectional view of human eye}
\label{fig:1}
\end{figure*}

\begin{figure*}
  \centerline{\includegraphics[scale=1.0]{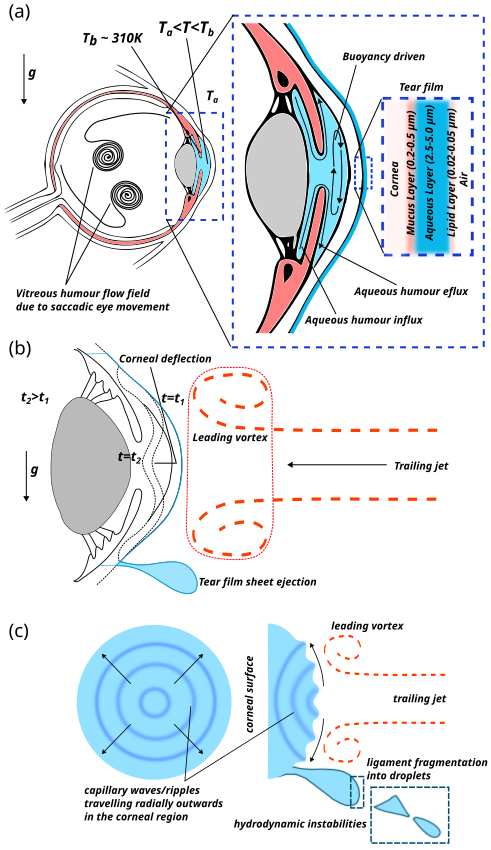}}
  \caption{(a) Schematic showing various flow fields that can exist in human eye under various conditions \cite{stocchino2007eye,braun2012dynamics,siggers2012fluid,fitt2006fluid}.
  (b) Deformation of the corneal surface during a typical NCT procedure and possible tear film ejection from watery eyes \cite{roy2021fluid}.
  (c) Schematic depicting the capillary waves/ripples propagating on the tear film surface on top of corneal surface. Hydrodynamic instabilities cause disintegration of tear sheet into droplets.}
\label{fig:2}
\end{figure*}

 The human eye, an essential part of the sensory nervous system, is one of our body's most complex organs and involves many physical, biochemical, and physiochemical processes \cite{albert2008albert,chalupa2004visual}.
 Human eye consists of various substances in different states of matter like solids, liquids, gels, colloids and soft materials to perform essential physiological processes and functions.
 For example, intraocular pressure (IOP) is maintained due to fluid pressure of 
 aqueous humour
 in the eye's anterior chamber.
Vitreous humour, another clear gel found inside the posterior chamber of our eyes helps to provide essential nutrients and maintains the shape of the eye. The corneal tear film is another important fluid responsible for several important physiological processes like protection against infection, remove free radicals, lubrication of the ocular surface \cite{dilly1994structure,pflugfelder2020biological}. Further, the tear film also provides a smooth optical surface for light refraction. 
 A thorough mechanistic understanding of the various processes and ophthalmologic conditions that occur inside our eyes is still elusive; and requires in-depth future investigation and analysis. 
 Ophthalmologic measurements like 
 IOP using NCT are based on several unexplored hydrodynamic processes and poses a challenge for scientists, engineers and medical professionals. 
 Understanding fluid mechanics in the context of the human eye is fascinating. It has significant implications for the accuracy of measuring devices like tonometers and hence is also important from a clinical perspective.\\
 \\
 The measurement of IOP using NCT is a transient hydrodynamic process and involves the interaction of an high speed air puff (velocity scale of the order of $5m/s$) with the cornea
 \cite{roy2021fluid}. 
 The critical interplay of external air pressure and intraocular pressure governs the corneal dynamics and its response. 
Further, corneal properties like elasticity, stiffness, and viscoelasticity are also important mechanical properties that play a crucial role in determining the corneal displacement profile and response time scale \cite{albert2008albert}.
 The amount of pressure required to flatten a specific area of the Cornea is used to calculate the IOP. Normal IOP ranges between 10 and 21 mmHg (millimeters of mercury) above atmospheric pressure but can vary between individuals and even throughout the day. Factors such as age \cite{tonnu2005influence}, genetics, and body position can all influence IOP levels. Higher IOP values are associated with an increased risk of developing glaucoma. 
 Fig. \ref{fig:1} shows a schematic representation of a human eye cross-section. The components labeled are the Cornea, Iris, Lens, Pupil, Aqueous Humour, Suspendory Ligaments, Cilia muscle, Eye muscle, Vitreous Humour, Sclera, Retina, and optic nerve. The Cornea is a transparent protective front eye part covering the anterior chamber, Iris, and pupil. The deformation and response of the Cornea on external loading are used to determine IOP. The important fluid elements present in our eye are the tear film, Aqueous Humour, and Vitreous Humour (refer \ref{fig:1} and \ref{fig:2} (a)).
 Several hydrodynamic mechanisms regulate IOP, including the production and outflow of aqueous humor in the anterior chamber (the region between the Iris and the Cornea) \cite{coakes1979method}. The ciliary body, a structure located behind the Iris and close to the ciliary muscle (refer \ref{fig:1}), is responsible for producing aqueous humor \cite{phd2002mechanism}. The fluid then flows through the pupil into the anterior chamber (fluid influx to the anterior chamber). Some fraction of it is drained from the eye via the trabecular meshwork and Schlemm's canal (fluid efflux from the anterior chamber). In some cases, the outflow of aqueous humor can become obstructed, leading to an increase in IOP \cite{johnson2006controls}. Some of the processes involving the dynamics of the tear film, aqueous humour, and vitreous humour under various conditions have been studied by mathematicians, physicists, and engineers over many years \cite{braun2012dynamics,stahl2012osmolality,tomlinson2005assessment,fitt2006fluid,siggers2012fluid}. However, similar comprehensive works in the context of non-contact tonometry is relatively sparse \cite{roy2021fluid}. Some recent works in the last decade have probed the fluid-solid interaction between the impinging air puff and the Cornea from a computational perspective using discretization schemes like finite volume, finite element, and arbitrary Eulerian-Lagrangian frameworks \cite{kling2014corneal,maklad2020simulation,simonini2016theoretical}.
Fig. \ref{fig:2} depicts various fluid mechanical phenomena schematically in the context of the human eye 
 (refer Fig. \ref{fig:2}(a)) and IOP measurements in general (refer Fig. \ref{fig:2}(b), (c)). For example, consider saccadic eye movements shown in Fig. \ref{fig:2}(a) that are responsible for transient flow field structures in the posterior chamber containing liquefied vitreous humour (ophthalmologic condition when the vitreous humour is liquefied or replaced after vitrectomy) \cite{stocchino2007eye}. The acceleration gravity is vertically downward depicted by $g$ refering to upright configuration of the eye.
Flow structures also exist in the anterior chamber containing aqueous humour. Flows in the aqueous humour are generated due to the combined effects of tear film evaporation; buoyancy-driven flow generated due to temperature gradients between the ambient ($T_a$) and the iris $T_b{\sim}310K>T_a$; and aqueous humour secretion by the ciliary bodies (refer Fig. \ref{fig:2}(a)). 
 These unique flow patterns in the vitreous and aqueous humour can be important for patient drug delivery. Fig. \ref{fig:2}(b) depicts some of the hydrodynamic processes and the air puff flow field during NCT. The air puff consists of a leading vortex followed by a trailing jet. Further the tear film sheet ejection is also depicted schematically initiated by the leading vortex. The trailing jet on impinging the corneal surface causes surface deflection. The corneal surface at various time instants ($t=t_1,{\:}t_2$) is shown schematically.
 Fig. \ref{fig:2}(c) depicts capillary waves radially propagating outwards on the surface of the tear film attached to the corneal surface. The tear film ejection as a thin sheet and disintegration into droplets through a series of hydrodynamic instabilities is also shown in Fig. \ref{fig:2}(c).
Elevated IOP values can lead to several ocular conditions, including glaucoma, which is characterized by damage to the optic nerve and loss of peripheral vision. In general, glaucoma is asymptomatic. However, specific symptoms such as eye pain, redness, and headaches can be associated with elevated IOP values \cite{sihota2000comparison}. If left untreated, high IOP levels can result in irreversible vision loss.
Treatment options for elevated IOP depend on the underlying cause and severity of the condition. Sometimes, lifestyle modifications such as exercise and a healthy diet can help lower IOP levels. Medications such as beta-blockers, prostaglandin analogs, and carbonic anhydrase inhibitors can also be used to lower IOP \cite{webers2010intraocular,liu2019efficacy,tejwani2020treatment}. In more severe cases, surgical procedures such as trabeculectomy \cite{kirwan2013trabeculectomy,jones2005recent,mills1981trabeculectomy}, laser trabeculoplasty \cite{damji1999selective,latina1998q,realini2008selective,latina2012selective}, or drainage implants \cite{schwartz2006glaucoma,netland1993glaucoma,lim1998glaucoma,prata1995vitro} may be necessary. 
Understanding the fluid mechanical phenomena in regulating IOP and identifying risk factors for elevated IOP can help prevent vision loss associated with ocular conditions. Regular monitoring of IOP is hence recommended, and NCT is one of the standard safe procedures.

When a puff of air from the tonometer nozzle is directed toward the cornea, a small indentation on the corneal surface is formed. The amount of indentation depends on various flow field parameters of the impinging air puff, such as velocity, pressure, and the distance between the air nozzle and the cornea. The corneal indentation required to estimate IOP will also depend on the anterior chamber's hydrodynamics. The flow field in the aqueous humour plays a critical role in determining the IOP. At a normal upright condition of the eye in an individual, at ambient temperatures lower than the average body temperature of $T_b{\sim}37^{\circ}C$, buoyancy effects cause convective flow fields to establish and generate velocity scales of the order of ${\sim}0.1mm/s$ in the Aqueous Humour (refer Fig. 2(a)) \cite{fitt2006fluid}. 
If the temperature of the ambient is higher than the body temperature $T_b$, there can be significant changes in the corneal tear film. Most importantly, the corneal tear film becomes thinner and becomes sticky as the concentration of mucus and lipids increases due to higher evaporation rates. Further, due to the heat transfer into the eye from the high temperature ambient, qualitatively the direction of the convection rolls reverses its direction in the aqueous humour as compared to the case where the environment temperature is lower than $T_b$.
The sudden interaction of the air puff with the cornea will set the aqueous humour into transient dynamics for a short time during the measurement process. Hence, the proper estimation of the IOP should be based on the dynamic nature of the flow field in the aqueous humour compared to hydrostatic pressure considered in the most estimation of IOP.
The indentation of the cornea causes a displacement of the aqueous humor. This displacement generates a series of waves that propagate through the fluid. The laws of continuum mechanics govern the propagation of these waves. The aqueous humor is an incompressible fluid, meaning its density cannot be changed significantly by applying external pressure. The incompressibility is due to a large bulk modulus of liquids in general and aqueous humor in specific. Therefore, any fluid displacement must be accommodated by a corresponding displacement elsewhere in the fluid. This displacement generates a pressure wave that propagates throughout the fluid. The speed of this wave depends on the bulk modulus and density of the fluid. During NCT, the pressure wave propagates through the aqueous humor and reaches the cornea. The cornea acts as a boundary condition for the wave, reflecting part of the wave into the aqueous humor and transmitting part into the eye. The transmitted wave propagates through the eye and is eventually dissipated as it encounters various structures in the eye. During the air puff ejection process, the air puff tonometer device also emits a collimated light beam that gets reflected from the cornea's surface to a measuring photocell sensor \cite{stamper2009becker}. The instrument is calibrated to measure the time delay for maximum reflection during the cornea's maximum deformation. The time delay of maximum reflection is calibrated with the force required to deform the cornea, hence measuring the IOP. The time delay is intricately related to the deformation of the cornea, which in turn is related to the various hydrodynamic interaction that occurs during the measurement process.
In a recent study, a group showed that a correlation exists between the tear film thickness and the IOP measured during NCT \cite{seol2019intraocular}. A higher value of the tear film thickness was shown to cause higher IOP measurements. Further, in our recent study, we also deciphered the mechanism of tear film destabilization that leads to droplet generation and possibly a pathogen transmission mechanism through the corneal tear film disintegration and aerosolization \cite{shetty2020quantitative,roy2021fluid}. The tear film disintegration process and aerosol generation were later confirmed computationally by a recent work of Zhou et al. \cite{zhou2022aerosol}. 
The mechanics of NCT are, therefore, complex and includes phenomena on various spatiotemporal scales. The measurement of the IOP depends on accurately estimating the applanation distance (distance between the tonometer nozzle exit and the cornea), flow field characteristics of the incoming jet, corneal elastic properties, and the aqueous humour. The understanding of these fluid-solid interactions is crucial for the optimization of the NCT instrument and the accuracy of the measurements. 
Further understanding the fundamental fluid-solid interactions during NCT can help medical practitioners deal with and prevent ophthalmologic complications like corneal perforation \cite{kim2008three}.
In this perspective, we provide some research directions that could improve our understanding of the phenomena at hand.

\section{Research directions}
Figure 3 depicts the time sequence of various fluid mechanical processes occurring during non-contact tonometry \cite{roy2021fluid} classified as different phases temporally. Some of the important phases of the non-contact tonometry process are the air puff ejection from the tonometer nozzle and the subsequent propagation of the air puff towards the cornea. As the air puff approaches the cornea, an initial tear sheet expansion occurs due to pressure gradients followed by the corneal deflection and propagation of capillary waves on the corneal tear film surface. The capillary waves in addition to the unsteady external air flow field expands the tear film into a 3D tear film sheet that breaks in droplets due to Rayleigh Taylor and Rayleigh Plateau instability \cite{roy2021fluid}. All the phases shown in Figure 3 could be its own research field. Some of the most important research directions are mentioned below.
 \begin{figure*}
  \centerline{\includegraphics[scale=0.8]{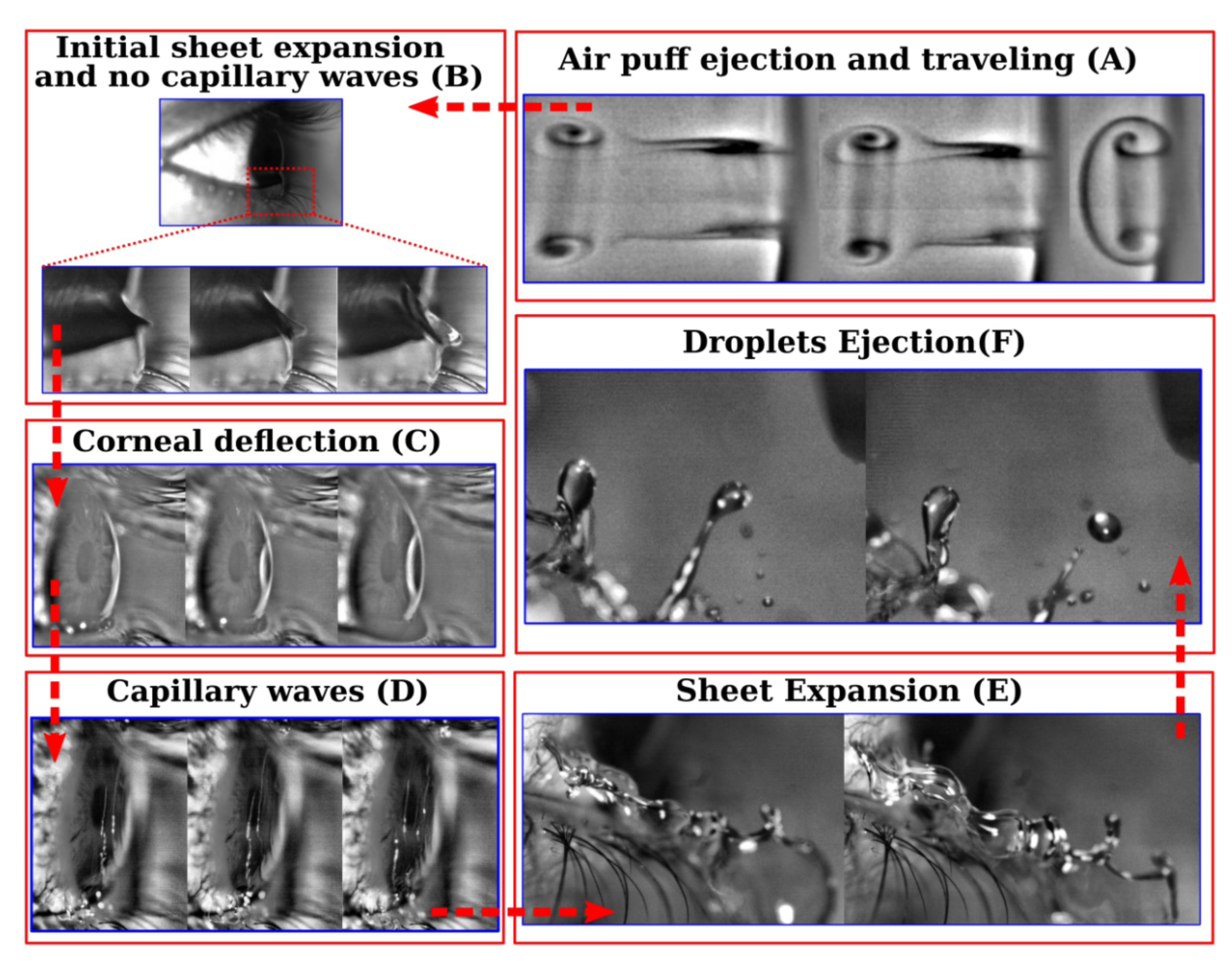}}
  \caption{Various fluid mechanical processes occuring during non-contact tonometry process labelled according to temporal evolution. The red dotted arrow depicts the order of events in time (A-F) \cite{roy2021fluid}.}
\label{fig:3}
\end{figure*}
\subsection{Understanding the incoming air puff}
 The air puff non-contact tonometer measures the force required to flatten a small area of the corneal surface with a puff of air. The air puff generated by the non-contact tonometer is an essential part of the measurement process. Understanding how its structure and flow field features affect the measurement process is crucial.
The air puff generated by the non-contact tonometer is a short air burst directed towards the eye \cite{stamper2009becker}. The force of the air puff is calibrated to be strong enough to deform a small area of the Cornea but not strong enough to cause any damage or discomfort. The air puff is typically generated by a small compressor that is built into the tonometer. When the measurement button is pressed, the compressor releases a brief burst of air that is directed toward the eye. The force of the air puff is calibrated based on the thickness and curvature of the Cornea. The tonometer measures the force required to flatten the Cornea and uses this measurement to calculate the intraocular pressure. While the air puff generated by the non-contact tonometer is not harmful, it can be startling to some patients. The sudden rush of air can cause the eye to blink, and some patients may feel a slight discomfort or pressure sensation. Proper optimized air puff design can help mitigate the discomfort. However, these effects are temporary and generally go away quickly. It is important to note that the air puff tonometer may not be suitable for all patients and age groups. Patients with certain eye conditions, such as corneal abnormalities or injuries, may not be able to tolerate the air puff test. Additionally, patients anxious or nervous about the test may find it challenging to keep their eyes open or remain still during the test. The air puff generated by the non-contact tonometer is essential to the intraocular pressure measurement process. While it may cause temporary discomfort or startle some patients, it is generally a safe and effective way to measure the pressure inside the eye. Although the measurement process of IOP using the air puff is being perfected by various device manufacturers, several aspects related to the structure of the incoming puff/jet are still unknown due to trade secrets and requires further experimental and theoretical investigations. In our recent work \cite{roy2021fluid}, using smoke flow visualization methods, we showed that the air puff essentially consists of a leading vortex followed by a trailing jet for the tonometer we used (refer to our previous work for details). Further, we could also measure the approximate air puff velocity using high-speed shadowgraphy \cite{settles2001schlieren} and particle tracking methods \cite{cowen1997hybrid}. Characterizing the incoming air puff is essential to properly understand the initial flow field features and characteristics subjected to the Cornea for IOP measurements. The incoming air puff could be characterized quantitatively with very high accuracy using high-speed particle imaging velocimetry \cite{adrian2011particle,willert1991digital}, particle tracking \cite{cowen1997hybrid} and shadowgraphy methods \cite{settles2001schlieren}. 
Further combining experimental with theoretical and computational studies will help scientists, engineers, and medical practitioners to understand the overall phenomena rigorously, which can help clinical practice.

\subsection{Fluid solid interaction between the air puff and the eye}
The air puff interacting with the eye deforms the cornea. The cornea's deformation is calibrated and directly related to the IOP values used by ophthalmologists. However, this process is not as simple as it sounds, as a complex fluid-solid interaction occurs during the measurement process. 
When the puff of air is directed at the cornea, it causes a deformation of the cornea's surface. This deformation creates a propagating wave, which travels through the cornea and the aqueous humor. The aqueous humor is a clear, watery fluid that fills the space between the cornea and the eye's lens \cite{phd2002mechanism,johnson2006controls}. The wave generated by the air puff propagates through the aqueous humor and interacts with the eye's lens. This interaction causes the lens to move slightly, which can affect the measurement of the IOP. The movement of the lens is known as the Ocular Response Analyzer (ORA) effect \cite{luce2005determining,kaushik2012ocular,terai2012identification,martinez2006ocular}.
Some tonometers use a dual-pulse system to compensate for the ORA effect. This system uses two puffs of air, one that is stronger than the other. The first puff is used to generate the propagating wave, and the second puff is used to measure the IOP. The difference between the two puffs allows for compensating the ORA effect. In addition to the ORA effect, a fluid-solid interaction occurs between the cornea and the aqueous humor. The fluid in the aqueous humor can act as a cushion and absorb some of the energy from the air puff. The energy loss can lead to an underestimation of the IOP. Some tonometers use a correction factor that considers the cornea's properties, such as its thickness and curvature, to compensate for the ORA effect. This correction factor helps to ensure that the measurement of the IOP is as accurate as possible \cite{ko2005varying,tonnu2005influence,cheng2006effect}. The complex fluid-solid interaction that occurs during non-contact tonometry measurements can affect the measurement of the IOP, and various methods are used to compensate for its effects. Eye care professionals need to be aware of these effects and use the appropriate techniques to obtain accurate measurements of the IOP.
The interaction between the air puff and the human eye can be mapped to a fluid-solid interaction (FSI) framework, which is quite a general approach. Some recent numerical works are aligned in this direction \cite{kling2014corneal,maklad2020simulation,simonini2016theoretical,maklad2023influence}. However, such computational models need to be corroborated with experimental measurements. Further, the computational model's initial conditions of the incoming jet should be taken from the kind of experimental/theoretical investigations mentioned in the previous section II. A. for determining the initial flow field structure of the incoming air puff/jet.
\subsection{Understanding various types of waves formed during the interaction}
The air puff interacting with the cornea and the tear film causes the formation of various types of waves like surface waves, capillary waves, shear waves, and lamb waves, to name a few. Understanding these waves is essential for accurate measurements of the IOP. The surface wave or S-wave is a type of wave that forms on the cornea and can travel through the entire eye during the measurement process. This wave is created by the sudden displacement of the air that is directed at the cornea. The surface wave is a small, high-frequency wave \cite{graff2012wave} that spreads outwards from the point where the air puff is directed at the cornea. High-frequency S waves being highly energetic, dissipates quickly.
In contrast, the capillary wave \cite{slavchov2021characterization} is a low-frequency wave that forms on the surface of the tear film on the cornea. Capillary wave is caused by the interaction of the air puff with the tear film surface, and surface tension plays a significant role in determining the wave characteristics. The capillary wave is slower and more long-lasting than the surface wave. Acoustic waves are another low-frequency wave that forms in the eye's aqueous humor \cite{de2001handbook}. This wave is created by the impact of the air puff on the cornea and travels through the aqueous humor towards the lens of the eye. Another low-frequency elastic wave that forms on the surface of the cornea and the entire spherical portion of the eye is Lamb wave \cite{zhang2013preliminary}.
These waves have different properties, such as frequency and amplitude, and can measure different aspects of the cornea and the eye. For example, the capillary wave can be used to measure the thickness of the tear film on the cornea. In contrast, the acoustic wave can be used to measure the depth of the eye's anterior chamber, and the Lamb waves could be used to measure corneal elasticity \cite{jin2020vivo}. Understanding these waves is essential for accurate measurements of the IOP and for diagnosing and managing various ocular conditions in general.
\subsection{Tear film interaction and dynamics, breakup and subsequent droplet formation mechanics}
The corneal tear film is a thin layer of fluid that covers the Cornea and is essential for maintaining the optical properties of the eye. The tear film comprises multiple fluid layers (mucus, aqueous, and lipids) \cite{braun2012dynamics}. 
Fig. \ref{fig:1.1}(a) schematically depicts the various layers present in the tear film. The mucus layer of thickness $0.2-0.5{\mu}m$ is the first layer just next to the Cornea, followed by a thick aqueous layer of thickness $2.5-5.0{\mu}m$. An extremely thin lipid layer of thickness $0.02-0.05{\mu}m$ exists just next to the aqueous layer and is the outermost layer exposed to the surrounding air. The tear film thickness varies between individuals ranging from dry eyes to watery eyes \cite{koh2015effect,cho2009dry,singh2019evaluation}. The tear film profile is almost uniform except at the base, where it meets the lower eyelid. The tear film thickness at the bottom-most point is the maximum.
The interaction between the air puff and the tear film (especially the bottom part) can lead to sheet formation and subsequent breakup in the case of watery eyes, affecting the accuracy of the IOP measurement. The deformation of the corneal surface due to the incoming air puff leads to various types of waves on the corneal surface, as discussed in section II.C above. These waves can cause disturbances in the tear film, leading to sheet formation followed by several hydrodynamic instabilities and resulting in tear sheet breakup. The breakup of the tear film can affect the measurement of the IOP in two ways. First, the breakup can decrease the force required to flatten the Cornea, which can further underestimate the IOP. Second, the breakup of the tear film can cause irregularities in the corneal surface, which can affect the accuracy of the measurement. Some tonometers use a double-shot technique to compensate for the breakup of the tear film, similar to that used for ORA compensation effects discussed in section II. B. This technique involves directing two air puffs at the Cornea, one stronger than the other. The first puff is used to generate the surface and capillary waves, and the second puff is used to measure the IOP. The difference between the two puffs allows for the compensation of the effects of the tear film breakup. In addition to the double-shot technique, other methods are used to reduce the effects of the tear film breakup. One method involves using a sodium hyaluronate solution \cite{gomes2004sodium} to coat the Cornea before the measurement. This solution helps to stabilize the tear film and reduce its breakup.
Another method involves using a video-based imaging system to monitor the tear film breakup during the measurement \cite{roy2021fluid}. This system can provide information on the stability and quality of the tear film, which can help to ensure the accuracy of the IOP measurement. 
The tear film is a crucial barrier in the eye that helps to protect against infection. During non-contact tonometry, the interaction between the air puff and the tear film can lead to its breakup, increasing the risk of pathogen transmission and infection. When the tear film breaks up, the underlying corneal surface is exposed, which can create a pathway for pathogen transmission. Pathogens such as bacteria and viruses can enter the eye through the exposed corneal surface, potentially leading to infections such as conjunctivitis \cite{azari2013conjunctivitis}, keratitis, or endophthalmitis \cite{henry2012infectious}. The risk of pathogen transmission is further increased when the tonometer probe comes in close proximity to the aerosols generated from the tear film resulting in the formation of fomites. Various precautions can be taken to reduce the risk of pathogen transmission during non-contact tonometry. These include ensuring that the tonometer probe is properly sterilized before and after each use. Disposable tonometer probes can also be used to prevent cross-contamination. In addition, proper hand hygiene should be observed before and after performing non-contact tonometry. Proper hygiene can reduce the risk of transferring pathogens from the hands to the tonometer probe or from the probe to the eye. 
The COVID-19 pandemic has highlighted the risks of pathogen transmission during non-contact tonometry \cite{roy2021fluid}. The interaction between the air puff and the tear film can lead to its breakup, increasing the risk of COVID-19 transmission and infection. The SARS-CoV-2 virus is primarily transmitted through respiratory droplets and aerosols. However, the virus has also been found in tears \cite{xia2020evaluation}, which suggests that transmission through the eyes is possible \cite{britt1991microaerosol,shetty2020quantitative,roy2021fluid}.
Various precautions can be taken to reduce the risk of COVID-19, like disease transmission during non-contact tonometry. These include using personal protective equipment (PPE) such as face masks, gloves, and eye protection.
It is also important to note that the risk of COVID-19 transmission during non-contact tonometry may vary depending on the prevalence of the virus in the community. In areas with high transmission rates, additional precautions may be necessary to reduce the risk of transmission. Therefore, understanding the tear film interaction and dynamics with subsequent droplet formation will be an important study to investigate the NCT process and its associated pathogen transmission routes and mechanisms. Several experimental techniques from fluid mechanics like interferometry \cite{hariharan2010basics}, shadowgraphy \cite{settles2001schlieren}, and laser Doppler anemometry \cite{durst1976principles}, to mention a few, could be used to study the tear film dynamics and its disintegration into droplets quantitatively.
\section{Conclusion}
In conclusion, we introduced a non-invasive eye procedure called non-contact tonometry highlighting various processes that occurs during IOP measurement. Further, we listed a series of research directions highlighting open and unexplored areas. Some of the research directions highlighted in this work include characterization and measurement of the incoming air puff, understanding the fluid-solid interaction problem between the impinging jet and the Cornea, understanding the various kinds of waves that form and travel on the eye during the NCT measurement process; and tear film breakup and subsequent droplet formation mechanisms at various spatiotemporal scales.  



\section*{Conflict of Interest}
The authors declare no conflict of interest.

\section*{Data Availability Statement}
Data sharing is not applicable to this article as no new data were created or analyzed in this study. 






\bibliography{aipsamp}

\end{document}